# POLITICAL AND LEGAL ASPECTS OF THE COVID-19 PANDEMIC IMPACT ON WORLD TRANSPORT SYSTEMS

**Alexey Gubin,**

*PhD in Law, Associate Professor*
*Law Institute of the Russian University of Transport, Russia, Moscow*

**Valeri Lipunov,**

*PhD in Law, Associate Professor*
*Law Institute of the Russian University of Transport, Russia, Moscow*

**Mattia Masolletti,**

*Lecturer, NUST University, Italy*

**Abstract**

The authors of the article analyze the impact of the global COVID-19 pandemic on the transport and logistics sector. The research is interdisciplinary in nature. The purpose of the study is to identify and briefly characterize new trends in the field of transport and cargo transportation in post-COVID conditions.

## 1. Introduction

The coronavirus pandemic has disrupted the usual ties between producers and consumers and has made serious changes in the business of logistics and transportation companies.

The spread of the coronavirus has dealt a serious blow to the global logistics and supply chain of raw materials and finished products. Thus, on April 7, 2020, major international organizations: the International Road Transport Union (IRTU) and the International Federation of Transport Workers (IFTW) had published an open letter to the governments of all countries with a request to support the transport industry in the context of the spread of COVID-19 [4]. The crisis caused an imbalance of cargo flows associated with changes in demand (see Fig.1), the suspension of production and the restrictions imposed. Due to this factor, state governments and international organizations should assign the highest priority to supporting the continuity and strength of supply chains.

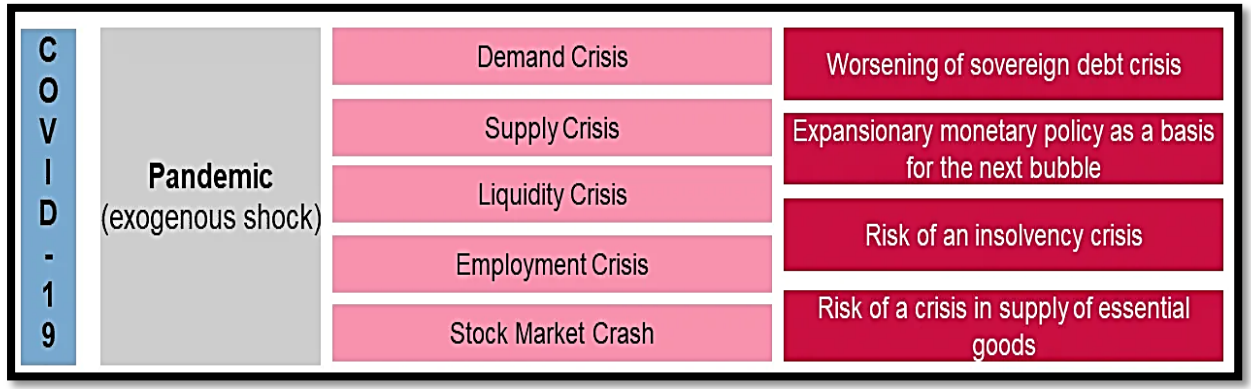

*Fig.1 Negative economic consequences of the COVID-19 pandemic*

Source: URL: **besthealthtourism.com**

## 2. Main part

Let us point out several main global transport and logistics trends:

- Reduction of cargo traffic on a global and local scale. The reasons are rather obvious: the closure of the borders of countries, the mass closure of retail outlets, the growth of the dollar, the isolation of the population, the decline in demand and purchasing power, as well as the state of fear and uncertainty among consumers. Many factories and factories around the world are closed for quarantine — there is nothing to transport and there is no one.
- The lack of simple, clear rules of the game in the conditions of quarantine for representatives of the transportation and logistics market.

Let's have a look at the current situation in various regions of the planet and start with China, which was the first to cope with the coronavirus pandemic.

There is a positive example of Asian countries, where the peak of the epidemic has passed and the cargo flow from countries to Europe is beginning to recover. For instance, 90% of China's production facilities have restored their work and are already sending cargo.

In China, all types of transportation were affected: air, sea, rail, and road. The usual multimodal [7] schemes were destroyed. Logistics companies had to urgently look for an alternative to the 'broken' links: for instance, to replace road transport within the provinces with rail. Due to problems with logistics, the most affected party could be the automotive industry, where the supply chain between auto parts production sites and assembly plants was under threat. However, it was possible to overcome logistical congestion [5].

In the conditions of the epidemic, rail transport has become the main tool in transportation. It was by trains that a significant part of China's anti-epidemic and medical cargo was transported. Railway groups of many Chinese provinces worked around the clock and helped factories, factories, construction and trading companies to return to work in a timely manner. In February 2020, Guangzhou restored railway communication with Russia. In May 2020, there had been a high demand for railway transportation from Asia.

It was also possible to solve problems on sea lines, despite the fact that maritime logistics is less elastic and requires more time to restore a normal rhythm. But even here, the operators had tried to respond flexibly to the situation by deploying emergency feeder services. Individual car companies had also worked effectively in difficult conditions.

The customs services of the Chinese provinces are also working effectively in the conditions of the epidemic, which have significantly reduced the time for processing priority cargo, opened 'green corridors' for anti-epidemic and medical cargo, as well as for raw materials and spare parts needed to restore production. Many customs offices have switched to the 'first release — then the end of customs procedures' mode [7]. It also helped to resume the work of enterprises and normalize foreign trade.

In general, cargo transportation is carried out, although the supply chain between China and Europe is still disrupted.

The EU economy is currently experiencing all the consequences of quarantine measures. The movement of freight transport has not been completely closed, but certain restrictions apply. In addition, transport companies have significantly fewer customers. The exponent of the number of cases in Europe is currently still growing.

In general, experts predict a drop in the European cargo transportation market by at least 40%.

The authorities of European countries introduce various preferences for the main players of the logistics market and, if companies have representative offices in Latvia, Germany, Italy, and other European countries, they will be able to reduce the tax rate and not pay rent during the crisis (you need to follow this information on the websites of the governments of the countries). If an organization needs to terminate contracts due to force majeure, it is necessary to obtain a document from the local authorities confirming the recognition of the epidemiological situation in the region as force majeure. For instance, in Lithuania, you can get such a certificate from the regional chambers of commerce and industry.

The main logistics trends in the EU include the following processes [2; 3; 6]:
- The volume of both intra-European and international traffic has decreased.
- Due to the self-isolation regime introduced in all European countries, the roads have become almost empty.
- The EU has lifted restrictions prohibiting the movement of freight transport on weekends.
- Rates within Europe are declining.

And now let's have a look at the current situation in Russia. At the moment, Russian logistics companies are experiencing difficult times. According to experts, only in Russia, the losses of the transport sector at the beginning of May 2020 had exceeded 230 billion rubles, and most of them had fallen on the aviation segment, which has practically stopped logistics operations. Warehouse areas are idle due to a decrease in the volume of cargo turnover. Railway operators, stevedores and road transport companies found themselves in a difficult situation. At the same time, there are also those to whom the pandemic has brought new prospects: logisticians note that the general trend is the shift of cargo flows to rail transport.

It is no secret that the two main flows of goods — the EU and China - were significantly reduced due to the pandemic. The markets for road, air and sea transportation are falling day by day, and there is no improvement yet. The government of the Russian Federation is introducing additional measures to support organizations, including logistics companies: tax holidays, deferrals on loan payments and related subsidies from banks, temporary cancellation of rent, and more.

Due to the economic consequences of the epidemic, imports of goods from the EU countries and, conversely, exports from Russia to the EU are decreasing. The strengthening of quarantine measures during customs clearance provokes delays and an increase in delivery times. As a result, the logistics chains of international transportation are changing and domestic traffic is growing.

According to InfraONE estimates, the losses of the infrastructure industries of the Russian Federation from the epidemic by September 1, 2021 will amount to approximately 507 billion rubles, of which almost 50% — 230.3 billion rubles-are losses of the transport industry. InfraONE evaluates, among other things, the effect of restrictions imposed to combat the spread of the coronavirus. As a result of the actions of the regions that have the right to review the terms of self-isolation of residents, the amount may decrease, 'but, most likely, this will be possible only in sparsely populated regions and will have little effect on the result'.

Airlines and airports suffered the most, which, according to experts, will lose about 270 billion rubles. The direction of international air transportation suffered the greatest losses. Airline flight schedules are constantly updated depending on the current epidemic and political situation. Due to the decrease in cargo traffic, many airlines operate on a charter schedule, applying tariffs with an increased coefficient of 2-3. The current rates are determined by the carriers at the time of booking and cannot be guaranteed for an extended period. It is worth noting that part of the cargo to Siberia and the Far East is sent instead of air delivery by highway transportation.

The railway transportation industry is experiencing a crisis. However, there are positive trends. The cargo flow from Asian countries is being restored. Against the background of the economic downturn, the 'Russian Railways' has taken unprecedented measures to stimulate transportation with discounts (for the transportation of coal, anthracite, etc.). Socially significant goods are also sent in covered wagons at a discount of up to 42.5% (by the way, the 'Russian Railways' notes an increase in food loading against the background of the epidemic: in March 2020 by 16.9% with an outstripping growth of individual items, including sugar — by 85.6%, vegetables — by 42.3%). The cyclical decline continues to develop in the operator segment — the slowdown in economic activity also affects the market for providing railcars.

In general, rail transportation to other destinations is complicated, and the rates have increased.

Despite the closure of borders, there is still a positive trend in the segment of Russian container transportation in the first quarter. But due to the complication of the situation in Europe, shipments from there to the East, as well as empty containers, have decreased.

Market participants note that the discount, despite the high share indicator, has little weight in the total cost of transportation. Historically, the container transportation segment has been more volatile due to its niche nature and a large focus on imports. In recent years, the level of containerization of exports has increased significantly, which can support a rapid recovery after the pandemic recedes, while imports and domestic transportation will remain under pressure from a weak ruble and the economy.

The peak of COVID-19 activity in Europe is the reason for the cancellation of the departure of ocean vessels from Southeast Asia, since it is impossible to process ship shipments in European ports. We are no longer talking about terms and tariffs. Often, companies are guided by the principle of 'just to deliver' [6].

Empty flights and the unstable situation at a number of border points significantly affect sea transportation:

- Ports are slower to accept and release cargo and with delays in customs clearance.
- Due to the drop in demand for transported products, transportation opportunities are reduced.
- There is an imbalance of free equipment (empty containers) by country: a shortage in some and a surplus in others. Ship-owners put ships on the dock until the demand for container transportation resumes.
- Many participants of flights are forced to stay at sea due to quarantine without the opportunity to 'go ashore'. The cargo 'hangs' in the sea. The sailors demand the unification of local quarantine measures within the country, the allocation of 'green corridors' for the implementation of personnel shifts.

The coronavirus epidemic has also affected road transport. Rates from the Russian Federation to the CIS countries (especially Kazakhstan) have risen by about 50%. Shipments to Moldova and Serbia are closed for the quarantine period, including for commercial vehicles. Priority or 'green corridor' for entry to all countries is given to cargo transport with food and medical goods. In general, imports and exports from Europe are carried out normally.

Queues at the borders of the EU countries have increased the transportation time. Due to additional sanitary checks at the borders, delays may occur, which negatively affects the delivery time of goods. Drivers are massively forced to comply with quarantine restrictions. The turnover of motor transport is falling, there is a shortage of goods due to the shutdown of production, a drop in consumer demand.

The intensification of the struggle for the client leads to the emergence of price dumping in the cargo transportation market, as the number of goods decreases, and transport is idle. Many companies will not be able to withstand long-term dumping.

In the near future, small and some medium-sized players will be forced to leave the logistics services market. Here, as in Darwin's theory, the strongest will survive. Those who have managed to save money over the past two 'fat' years and have not burdened themselves with significant obligations will now definitely look at the opportunity to buy something, increase the fleet and market share. There will be a series of bankruptcies, mergers and acquisitions.

Players will begin to unite in communities to share each other's services. Logistics and service companies are beginning to collaborate, develop unique comprehensive offers for customers and, as a result, strengthen their joint positions by combining their services.

90% of logistics companies refuse to update their fleet due to the growth of the exchange rate and the pandemic. This means that the fleet of cars will become obsolete, services related to the repair and maintenance of fleets will be in demand.

There is a tendency to reduce the batches of delivered goods and an increase in the number of combined cargo sent. Significant restrictions on air transportation will 'transfer' part of the demand from cargo owners for the transportation of combined cargo. The development of outsourcing will teach market players to 'fill' vehicles and group shipments on mutually beneficial terms. There is a particularly high demand for combined cargo from European countries.

Optimization and digitalization have been discussed for a long time, but only a few people decided on real and fundamental changes in the approach [8]. Many companies during the COVID-19 pandemic decided to transfer all their work to 'new rails'. The IT revolution has begun in logistics. It is worth noting the use of IT platforms for logistics companies in order to exchange tariffs and rates.

A highly automated logistics chain is now in great demand. Cargo owners need a full range of services with access mode from their mobile device. The client gets the opportunity to order transportation on a digitalized logistics platform by clicking on a button in the mobile application. It is such systems that will be in demand in the future. For example, for individuals, the Russian Post has launched a new service for sending parcels by phone number — the function is available to all users of the logistics operator's mobile application. The sender only needs to enter the recipient's phone number or select it from the phone book. It is also worth noting the high service of the companies 'Dostavista', 'Peshkariki', 'Garantbox', 'Take'n'go', 'Boxberry', 'Scooter', 'Yandex Delivery'.

Until recently, most transportation was directed to the export or import of goods, neglecting domestic markets. The crisis gave a powerful impetus to the development of the domestic product, the development of production within the country. A significant decrease in cargo flows from other Asian countries and the threat of closing borders with China in the event of the next wave of the pandemic lead to the fact that some of the resources, goods, products that were previously purchased there, manufacturers will try to produce in their own country. For example, the top 5 goods that are supplied to Russia from China — smartphones, garlic, professional sports equipment, chemical fertilizers, clothing and shoes — can be produced inside the Russian Federation. Now I want to believe that this is a chance for Russia to close the logistics chain on itself. There will be an increase in domestic production due to the closure of borders and, as a result, the development of internal logistics, reaching a new level of quality.

The trend of outsourcing non-core processes and services will gain great momentum. Although this trend has been actively developing since the 2000s, now it carries not only the possibility of saving the budget, but also significant time savings.

## 3. Conclusion

Thus, in connection with the global crisis provoked by the COVID-19 pandemic, the logistics industry is in urgent need of support. The quarantine measures taken to suppress the coronavirus epidemic have led to the congestion of most airports and sea terminals and, as a result, the violation [1] of the conditions and terms of cargo delivery.

Transport logistics incurs losses. A number of experts say that, for example, many airlines have a safety margin of no more than two months. It will not be easy for ship-owners, but they have already learned since 2008 to work with varying success in changing conditions. It is most difficult for road transport companies: there is practically no safety margin. Everything depends on financial partners, namely leasing companies, banks and car manufacturers that lend to car enterprises. The maximum period that they can withstand is up to six months. Then another wave of bankruptcies may come.

Railway carriers are often mentioned among those to whom the pandemic has given new opportunities. Railway transport operators do not respond to changes in demand by increasing rates, which is why this type of transport seems to be the most reliable and efficient when transporting goods between Russia, the EU and China in the current conditions. Railway transport, as a rule, is under the jurisdiction of the state — it will be supported. In addition, the railway will become one of the main logistics channels in the next few years to ensure uninterrupted trade between the Russian Federation, China and Europe and the delivery of anti-epidemic drugs.

State support measures are being actively implemented to overcome the crisis in the logistics industry. As a rule, support is provided to the state transport sector (railways, aviation). State support can be provided to the commercial sector through the introduction of tax holidays, the abolition or reduction of road tolls, as well as financial assistance to companies that will suffer large losses due to the pandemic. If we talk about road transportation, it will be significant to help reduce the cost of spare parts for vehicles, for example, by reducing import duties or the VAT rate for this group of goods. Very effective measures could also be: a moratorium on fines (except for fines regulating road safety), the abolition of toll collection on federal highways, tax exemption for the most unprotected carriers — sole proprietors with one or two cars, suspension of lease payments without fines.

Overcoming the current crisis is an unprecedented challenge for the management team of all logistics players. It is necessary to promptly implement all available state support measures and develop a further action plan. This will require the resources of a whole team, including lawyers, financiers, and economists. At the same time, now is the time to review partnerships with contractors, deadlines and obligations under contracts, carefully work with working capital and liquidity.

**References**

[1] Afanasyeva, O. et al. (2019) The condition and trends of crimes committed in public places // RUSSIAN JOURNAL OF CRIMINOLOGY. - Volume13. Issue 6. - pp. 895-908. – DOI: 10.17150/2500-4255.2019.13(6).895-908

[2] Ahlfeldt, G., & Feddersen, A. (2010). From the periphery to the core: The economic adjustment of high-speed rail. Institute of Economics of Barcelona, Working Paper No. 38, October 2013, accessed: http://dialnet.unirioja.es/servlet/articulo?codigo=3284478

[3] Bagreeva E. G., Zemlin A. I., Shamsunov S. K., Blankov A. S. (2021). On the classification of environmental safety risks of the transport complex: legal and organizational aspects // Turismo-estudos E Praticas, Brazil.

[4] Besinovich, N. (2021). Assessment of the impact of COVID-19 on the capacity of railway networks // European Journal of Transport and Infrastructure Research (EJTIR) / https://doi.org/10.18757/ejtir.2021.21.1.4939

[5] Boehm, M., Arnz, M. and Winter, J. (2021). The potential of high-speed rail transport in Europe: how is the transition from road to rail transport possible for low-density and high-cost goods? Euro. Transp. Res. Rev. 13, 4. - https://doi.org/10.1186/s12544-020-00453-3

[6] Braby D., Andersson E. (2008). Review of some cases of high-speed train derailments: means of minimizing the consequences based on empirical

observations. Department of Railway Vehicles, Department of Aeronautical and Automotive Engineering, Royal Institute of Technology (KTH). – Stockholm.

[7] Cheng, Yu. (2010). High-speed railways in Taiwan: new experiences and challenges for future development // Transport Policy, Volume 17, No. 2, pp. 51-63.

[8] Dmitriev, A.V. (2018) Digitalization of transport and logistics services based on the use of augmented reality technology. Bulletin of SUSU. Vol. 12. pp. 169-178.